\documentclass[sigconf]{acmart}

\AtBeginDocument{%
  \providecommand\BibTeX{{%
    \normalfont B\kern-0.5em{\scshape i\kern-0.25em b}\kern-0.8em\TeX}}}

\copyrightyear{2021}
\acmYear{2021}
\setcopyright{cc}
\setcctype{by}
\acmConference[Asian CHI Symposium 2021 ]{Asian CHI Symposium 2021 }{Nov 19-21, 2021}{Yokohama, Japan}
\acmBooktitle{Asian CHI Symposium 2021 (Asian CHI Symposium 2021 ), Nov 19-21, 2021, Yokohama, Japan}
\acmDOI{10.1145/3429360.3468186}
\acmISBN{978-1-4503-8203-8/21/05}

\begin{document}

\title[Refresher Training through Quiz App for Community Healthcare Workers in India]
{Refresher Training through Quiz App for capacity building of Community Healthcare Workers or Anganwadi Workers in India}

\author{Arka Majhi}
\email{arka.majhi@iitb.ac.in}
\affiliation{
  \institution{Indian Institute of Technology Bombay}
  \streetaddress{Powai}
  \city{Mumbai}
  \state{Maharashtra}
  \country{India}
  \postcode{400076}
}
\orcid{https://orcid.org/0000-0002-5057-1878}

\author{Satish B. Agnihotri}
\email{sbagnihotri@iitb.ac.in}
\affiliation{
  \institution{Indian Institute of Technology Bombay}
  \streetaddress{Powai}
  \city{Mumbai}
  \state{Maharashtra}
  \country{India}
  \postcode{400076}
}
\orcid{https://orcid.org/0000-0002-0703-3185}

\author{Aparajita Mondal}
\email{aparajita.mondal@iitb.ac.in}
\affiliation{
  \institution{Indian Institute of Technology Bombay}
  \streetaddress{Powai}
  \city{Mumbai}
  \state{Maharashtra}
  \country{India}
  \postcode{400076}
}
\orcid{https://orcid.org/0000-0003-4609-2249}

\renewcommand{\shortauthors}{Arka Majhi, Satish B. Agnihotri, and Aparajita Mondal}

\begin{abstract}
High and persistent child malnutrition levels with tardy reduction, seen in successive health surveys, continue to be a matter of concern in India, drawing attention to the need to revamp the four-decade-old Government program, Integrated Child Development Scheme (ICDS). ICDS field functionaries or Anganwadi Workers' (AWWs) capacity deficit was identified as a significant factor affecting ICDS's effectiveness. Considering rising numbers, over 1.4 million AWWs, and continuously advancing knowledge of community healthcare, conventional training pedagogy is ineffective in building and updating AWWs and their supervisors' capacity, which calls for rethinking, using the ICT approach. Over 6 lakh AWWs in India were smartphone equipped by 2020. An android based quiz app was designed, following AWWs training modules' content and need assessment results. The study investigates the quiz app's effectiveness and compares it with conventional classroom instruction, with a group of AWWs, and discusses ways to make it an adequate substitute.
\end{abstract}



\maketitle

\section{Introduction}
Healthcare systems worldwide are struggling with a shortage of Community Health Workers (CHWs) or trained health workers, especially in low-and-middle-income countries (LMICs), which is estimated to reach a deficit of 80 million by 2030 \citep{WHO2016}. It remains a challenge for LMICs to train or educate CHWs ensuring the most vulnerable population's good health and well-being \citep{WHO2016}. High and persistent child malnutrition levels with tardy reduction, seen in successive health surveys, continue to be a matter of concern in India, drawing attention to the need to revamp the more than 45-year-old flagship program of the Government of India, the Integrated Child Development Services (ICDS), under the Ministry of Women and Child Development (MWCD), focusing on children (below 6 years), pregnant women, and lactating mothers \citep{Rao2018, Gragnolati2006}. ICDS offers a comprehensive package of services, including supplementary nutrition, immunization, health check-up, referral services, education on health and nutrition, and pre-school education to break the deadly loop of malnutrition, morbidity, and reduce learning capacity and mortality. Iron and folic acid (IFA) tablets and vitamin A dozes are administered to prevent iron deficiency anemia and Vitamin A deficiency. These services are offered through a part-time honorary field functionary of the ICDS program, residing in the same locality called Anganwadi worker (AWW). AWW is assisted by an honorary helper called Anganwadi Helper (AWH), a local resident. This duo serves at a village centre called Anganwadi Centre (AWC), serving a population of 400-1000 in rural and urban areas and 300-800 in tribal and hilly areas.

Finance Minister of India, while presenting the Union Budget 2020-21, announced that over 6 lakh AWWs are equipped with smartphones and can upload the nutritional status of more than 10 crore households. A majority of it got entered through a mHealth platform, CommCare, a mobile application developed by Dimagi, USA \citep{Chanani2016} commonly known as ICDS-CAS \citep{Nimmagadda2019}. It helped reduce the human errors in classifying the grade of child malnutrition after measuring the height, weight, and MUAC, which was traditionally done through printed tables and now aided by algorithms in the app. With the improvement of technology and penetration of smartphones to cities, rural and tribal villages of India, access to digital training systems can also benefit the AWWs and the healthcare system, rendering reduced malnutrition and infant mortality rates.

The purpose of this paper was to design a smartphone-based refresher training tool, a quiz app for AWWs in India, and get early feedback on the app design and usage from AWWs by conducting a pilot testing through field study. Our aims were to 1) Build a quiz-based app with contents from training modules of AWWs 2) Compare the effectiveness of the refresher training through the designed quiz app with the traditional classroom-based training. We hypothesized that there would be deviations of knowledge retention through quiz app-based learning pedagogy compared to classroom-based refresher training. We based our research on how narrative or storytelling-based quizzing might support AWWs knowledge, skill, and capacity building. We also highlight the challenges faced by the AWWs during the learning process and the role smartphone-based quiz app intervention can play. Through qualitative and quantitative analysis, we found that the AWWs achieved significant knowledge gains and retention by playing the Anganwadi Quiz app.

\section{Literature Review}
Mobile Health or mHealth is defined as "medical and public health practice supported by mobile devices" \citep{WHO2011}, thus having the potential to overcome the limitations of the traditional methods of receiving child and maternal healthcare through job cards or printed paper brochures. mHealth is a means to disseminate healthcare information cost-effectively and personalized to socioeconomic and cultural needs to fit the context. The rise in smartphone ownership among healthcare workers in India opens up an opportunity to utilize mHealth to bridge the gap of healthcare access and engagement with the beneficiaries of the target population. 

Researchers from developing countries have provided suggestions for designing effective mHealth interventions \citep{Kumar2015}. These include communicating with the community through radio and text messages (both automatic and human aided). Studies have been conducted to motivate health workers and increase their knowledge by deploying short videos on mobile phones \citep{Ramachandran2010}. Another study \citep{Kumar2015} from India found that watching short films on mother and child care empowered Health Workers and mothers to manage pregnancies proactively. Shah. et al. \citep{Shah2017} from India did a similar study on an incentive-based approach for training AWWs through mobile-based videos. The AWWs' feature phones were loaded, with videos related to mother and child healthcare. A few questions with their correct answers related to the content of the video were appended to the video.  After watching these videos, AWWs were supposed to call on a toll-free number, where they would be quizzed on the same questions as shown in a particular topic video.  AWWs were awarded cellular talk time as an incentive for answering correctly. Another study \citep{Perez2020} from India, found that Mobile Vaani, an IVR based mHealth solution to improve knowledge on immunization have shown significant results to improve healthcare knowledge, if the content is customized to meet the needs of less-literate users. But none of these interventions in India and other developing countries have tried serious games or game-based learning as a pedagogy for mHealth intervention.

To make the learning experience of the AWWs more engaging for AWWs, there is a need of a platform where the teaching, learning and evaluation happens under the same umbrella accompanied by some thrill and fun of mobile gaming. It should also be scalable for a large number of AWWs, remotely configurable and progress/usage monitor able. Also the knowledge base should be easily update-able. A self-paced refresher training for the AWWs have become a necessity, for all AWWs in India. With the improvement of technology and penetration of smartphones to cities as well as rural and tribal villages of India, the access to digital training systems, could benefit the AWWs and the healthcare system to a large extent, rendering reduced malnutrition and infant mortality rate, countrywide.



\section{Design Overview}

Using quizzes to engage students is practiced in academia. Trainers are incorporating quizzes into their curriculum, trying to create a fun and engaging learning environment. The purpose of the quiz app is to assess and give refresher training to the AWWs and monitor the learning over a period through a modeled quiz. We adopted the Incremental Learning Approach (ILA) modules as the learning content, as the AWW trainers also used them to train AWWs. ILA modules are 21 learning modules and 21 takeaway modules prepared under National Nutrition Mission (NNM) (which is now POSHAN Abhiyan) to provide the necessary job training to the AWWs through classroom training and e-ILA through digital platforms \citep{ILA2021}. The topics of ILA module mainly focuses on the first 1000 days of life of a child, starting from conception, pregnancy, antenatal care, delivery, postnatal care unto a child at 2 years. It also focuses on adolescent and maternal health. A sample module of takeaways from Anemia module is attached to the appendix section, for reference. We categorized the ILA modules into 6 broad topics: anemia, pregnancy, breastfeeding, complementary feeding, identifying children with illness, and observing growth in children. The authors selected the essential questions which covered the content of each module. Then, the authors prepared storyline narratives from those questions.

Every time an AWW plays the quiz app, a story narration opens up. An audiovisual onboarding shows that she will be visiting a home in a village where she listens to a medical condition of mother or child or both. The mother/ daughter/ father/ child narrates the symptoms/ problems. AWW has to identify the medical condition the mother/child has and answer the health issue questions related to that topic. These questions include the causes, remedies, medicine dosage, side effects, and anticipations if it aggravates suddenly. With each right answer, the player gets a star badge as a prize. The correct information/ knowledge prompt is shown on both winning and losing, where the reinforced learning happens. It also reveals more information about the topic for better clarification and shows its relevance to other diseases. It creates interest by showing exciting facts and figures about the issue.  The scoreboard and leaderboard show up on completing a quiz round. The AWW can revisit the previous questions by playing the round again and correctly answer them, indicating the last round’s refresh of knowledge. The objective of the quiz is not only to judge the AWW’s knowledge retention but also to refresh their knowledge.




\section{Methods}

\subsection{Context}
Dharavi is a slum in Mumbai city in India. There are 300 AWCs in total, spread over slightly larger than 2.1 square kilometers with a population about a million. Considering this to be a good representation of urban AWWs, Dharavi was chosen as the cluster for pilot testing the quiz app. Permission for conducting the study with the AWWs, in the cluster of Dharavi, was granted from the CDPO of Dharavi, Mrs. Shobha Shelar, and assisted by the AWW supervisor, Mrs. Vijayata Vyapari. The AWWs gathered for their scheduled training in 5 batches alternatively, at Chota Sion Hospital in Dharavi.

\subsection{Participants and Sampling}
Among the Anganwadi Workers (AWWs), the ones who were willing to take part in the study, (n=154) were chosen as sample. The selected AWWs were aged between 31-63 years. Their field experiences varied from 2-34 years. The study was conducted from May to June, 2018. 60\% of the AWWs were educated till ninth standard. The inclusion criteria was that all AWWs should have an Android based smartphone and should be fluent in reading, speaking and writing in Hindi, as the app was also designed in Hindi. Considering the majority of Indian population being native Hindi speaker, this decision was taken. A written consent was taken from all AWWs participating in the study. 

\subsection{Study Design}
The study was conducted in 3 phases: 1) Written Pre-test with questionnaires printed on paper 2) Learning through traditional method or through the quiz app 3) Written Post-test with questionnaires printed on paper. The AWWs were divided into two almost equal major groups (Group 1 and Group 2). Both of these groups were intervention groups. There were no control group. One of the intervention group (Group 1) (n=76) was tested with traditional method by distributing reading material on Anemia, and encouraged to read it for a week. Another intervention group (Group 2) (n=78) was trained through the designed quiz game on anemia with prompts for learning, installed in their smartphones and asked to play for a week, whenever they get time. 

\begin{figure}[h]
  \centering
  \includegraphics[width=\linewidth]{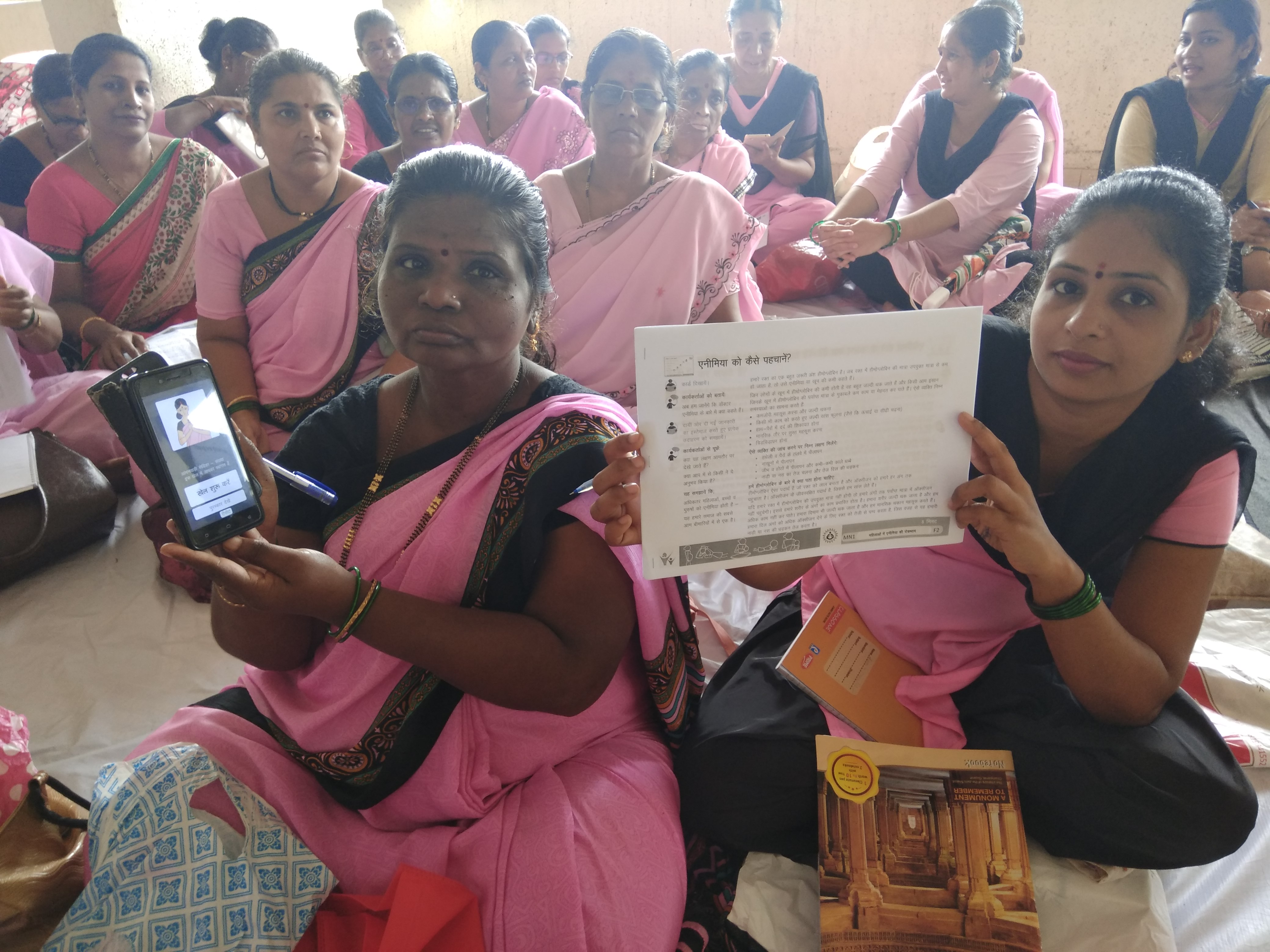}
  \caption{Anganwadi Workers - Intervention group 1 (right) and Intervention group 2 (Left)}
  \Description{One Anganwadi Worker at right holding the ILA brochure of Anemia and another Anganwadi Worker at left holding smartphone. A bunch of other Anganwadi workers sitting behind them}
\end{figure}

Pre and post test, containing 10 Multiple Choice Questions with some single select and some multiple choice answers. All the questions were equally difficult and based on the contents of the booklet and app. The questions of Test 1 were asked in a different way in Test 2. It was made sure that the content of the questions of the tests were included in both of the traditional reading material and the designed quiz game. For each intervention group (Group 1 and 2), two subgroups were formed (Group 1A\&1B and 2A\&2B). Group 1A (n=39) was given Test1, as a pre test and Test 2, as a post test and group 1B (n=37) was given Test2, as a pre test and Test 1, as a post test. The same procedure was repeated for Groups 2A (n=40) and 2B (n=38) also.  

The pre and post test questionnaires from the above mentioned 4 groups were evaluated by the authors. Each right and complete answers were awarded 2 marks, while the correct but incomplete ones were awarded 1 marks. The wrong answers scored 0 marks. It was expected that participants will score better in the post test than a pre test, irrespective of the interventions in between. But this test was done to compare the differences,  between both the intervention groups (compare between groups), between pre and the post test (compare within groups), to compare the learn-ability of both the methods (text book based and quiz app based).

\subsection{Implementation}
During the training sessions, the reading materials were distributed to the intervention group 1. The application was installed in the smartphones of AWWs of the intervention group 2. Group 2 was briefed about the game-play and encouraged to play whenever they get time over a period of a week. Quiz App playing sessions shown in the Figure 2 were recorded for further analysis. On the other hand, the intervention group 1 was trained through classroom training following ILA modules or brochures and encouraged to read it when they get time over a period of a week.

\begin{figure}[h]
  \centering
  \includegraphics[width=\linewidth]{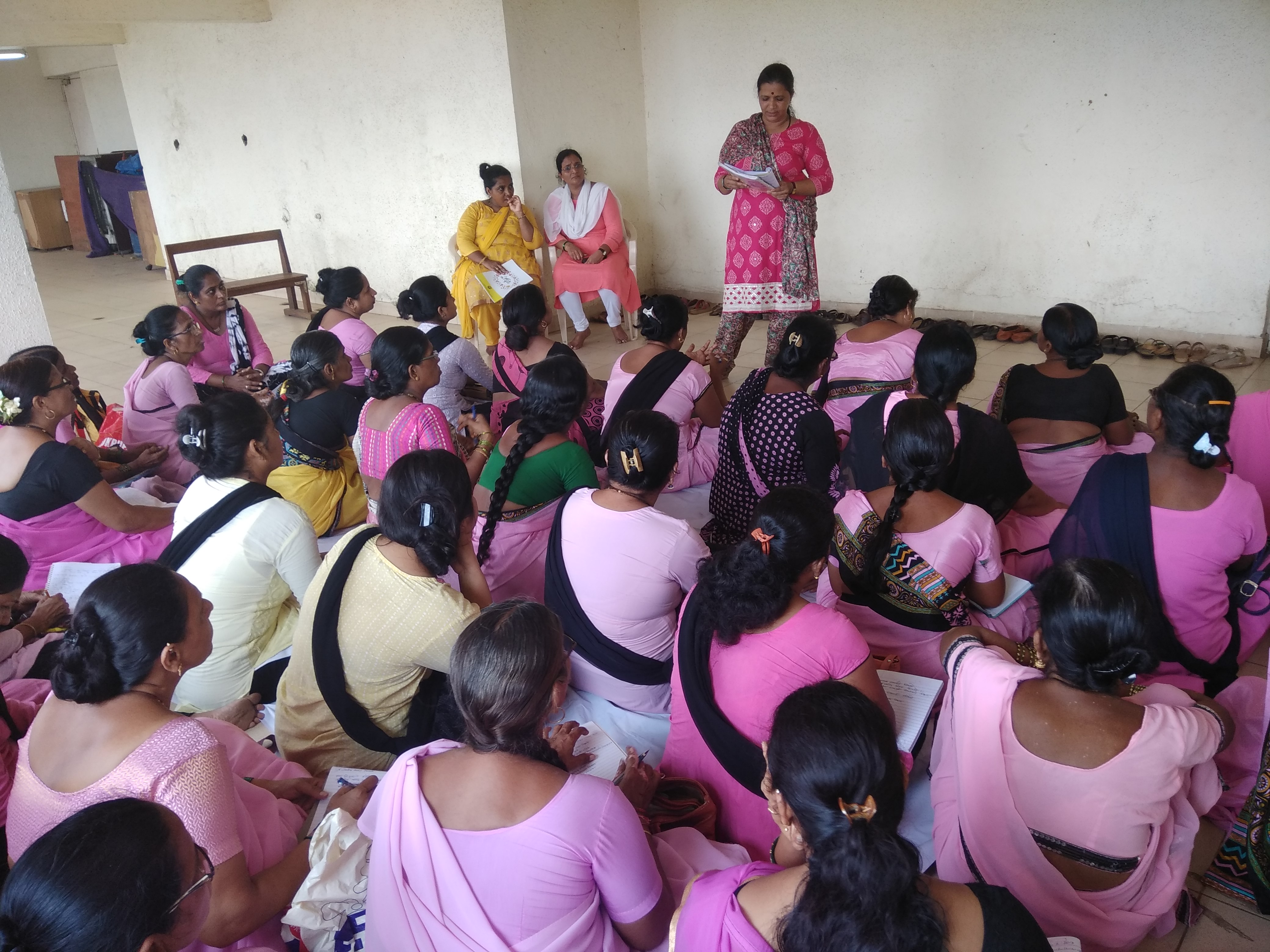}
  \caption{Anganwadi Workers being trained by trainers and their supervisor in a classroom setting}
  \Description{A bunch of Anganwadi Workers sit on the floor in rows and a trainer stands in the front row facing the Anganwadi Workers. The trainer reads out the reading material and conduct the training assisted by the supervisor.}
\end{figure}

\begin{figure}[h]
\centering
  \includegraphics[width=\linewidth]{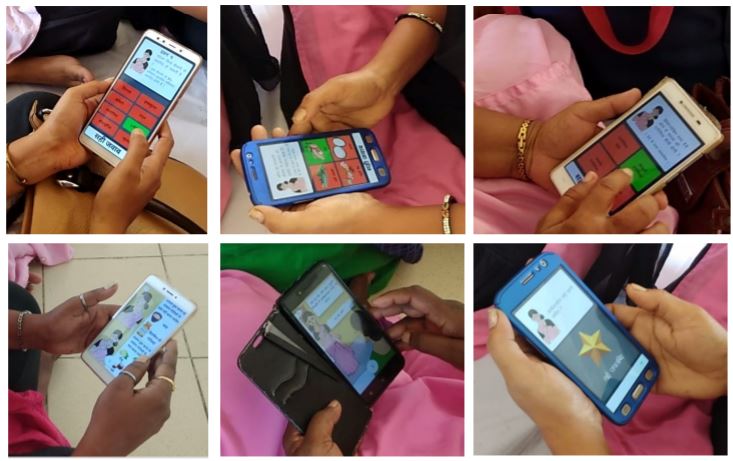}
  \label{fig:PlayingQuizApp}
  \caption{Anganwadi Workers playing the quiz app on their smartphones}
  \Description{A bunch of Anganwadi workers playing the quiz app on their smartphones}
\end{figure}

Pre-test and post-test were written tests on questionnaires printed on paper similar to the questions asked in the quiz app or in the printed brochures. These question papers were distributed to the AWWs and given 15 minutes to fill and return back the questionnaire. AWWs are supposed to write their name and details and answer 10 questions by putting a tick on Multiple choice questions with a single answer.

\begin{figure}[t]
  \centering
  \includegraphics[width=\linewidth]{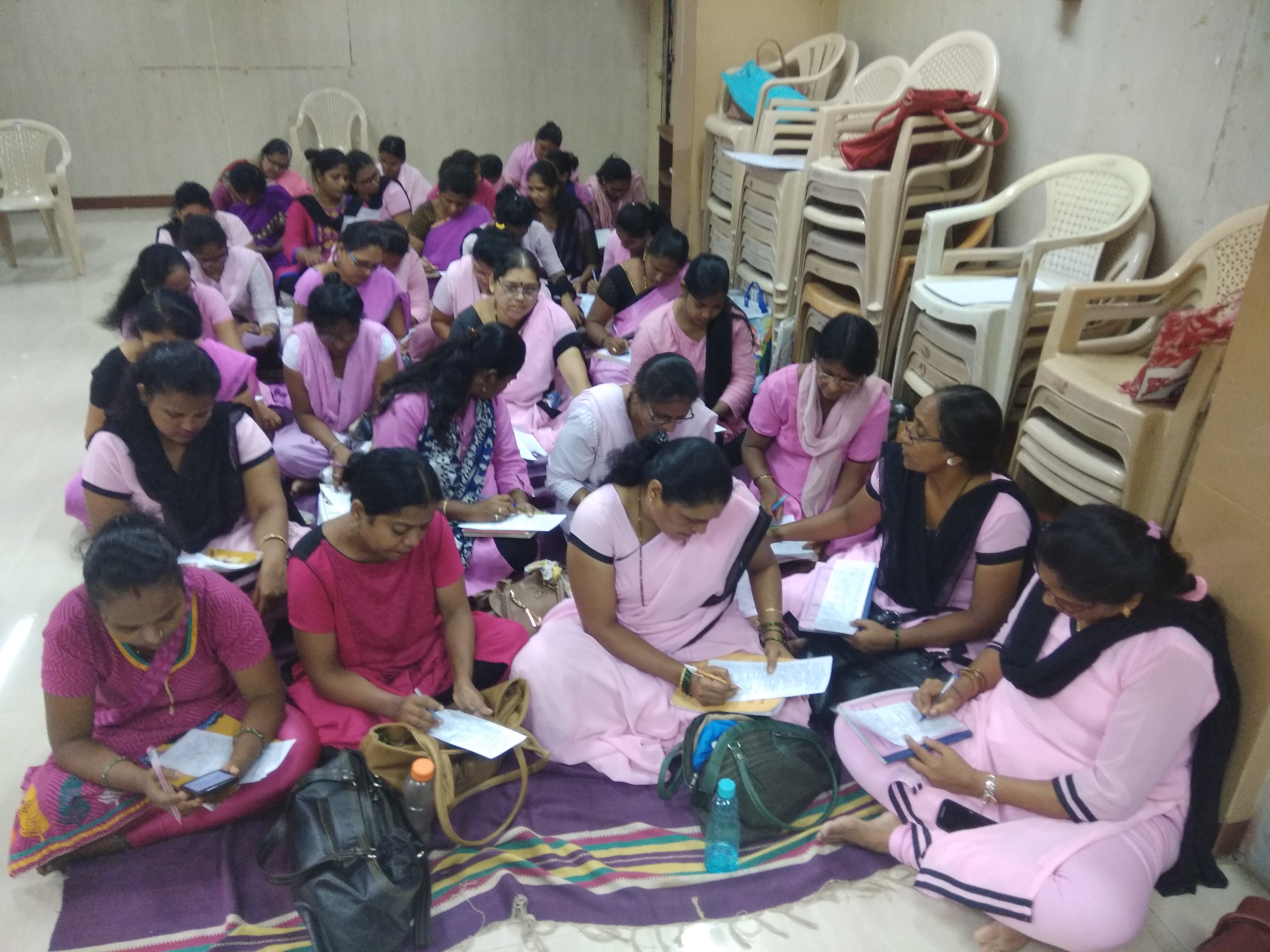}
  \caption{Anganwadi Workers participating in Pre-test and Post-test}
  \Description{A bunch of Anganwadi workers sitting on the floor and participating in written Pre-test and Post-test}
\end{figure}

\section{Results}
The core objective of the study was to measure the effectiveness and knowledge retention after the AWWs' learning from the quiz app and compare it to that of traditional method of classroom learning through printed brochures. The mean difference of scores of pre-test and Post-test were compared. Knowledge gained by AWWs were assessed to compare effectiveness of the learning interventions.

\subsection{Data Analysis: Comparing pre-test and post-test}
On one hand there were 78 participants in the intervention group 2 using app for learning. The mean pre-test score of the intervention group 2 was 5.92 and the mean post-test score was 8.56 leading to a difference of 2.64. On the other hand there were 76 participants in the intervention group 1 using app for learning. The mean pre-test score of the intervention group 1 was 5.60 and the mean post-test score was 7.44 leading to a difference of 1.84. The differences between the mean scores in pre-test and post-test of both groups clearly indicates that after both the interventions, the AWWs gave more correct answers. The differences between pre-test and post-test scores are summarized in Table \ref{tab:diff}

\begin{figure}[h]
\centering
  \includegraphics[width=\linewidth]{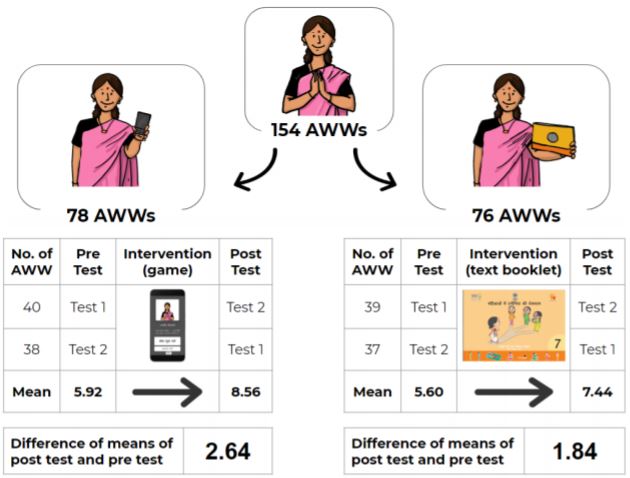}
  \label{fig:PlayTesting}
  \caption{Study Design and Pre-Test \& Post-Test results}
  \Description{154 AWWs broken up into 78 and 76 AWWs with 2 tables of pre-test and post-test results}
\end{figure}

\begin{table}
  \caption{Summary of means of pre-test and post-test scores and their difference. Learning through Printed brochure (Intervention Group 1 or IG 1). Learning through Quiz App (Intervention Group 2 or IG 2)}
  \label{tab:diff}
  \begin{tabular}{cccc}
    \toprule
    Intervention&Pre-test score&Post-test score&Difference\\
    \midrule
    IG 1&5.60&7.44&1.84\\
    IG 2&5.92&8.56&2.64\\
  \bottomrule
\end{tabular}
\end{table}

\subsection{Comparison of interventions}
In order to find the effectiveness of the interventions, the difference (within group) in scores of the pre test and post test between (between groups) both the intervention groups were compared using quantitative data analysis. A paired two-sample t-test, assuming unequal variance was performed with the collected data. Null and Alternate Hypothesis for the test was considered as shown in Table \ref{tab:hypothesis}. The results of the t-test are summarized in Table \ref{tab:t-test}. Since, the P-value (two-tail) derived from t-test was '0.000000014' which is lesser than '0.05', the null hypothesis (\(H_0\)) which states that, there is no significant difference between the intervention groups 1 and 2 (between groups) in terms of the difference of means of pre-test and post-test scores (within group), thus the null hypothesis was rejected. This means there was a difference between the effectiveness of pedagogy involved in both the intervention groups. Correct answers given by the AWWs in Group 2 with quiz app intervention were greater than correct answers by Group 1 with traditional method of learning from printed brochures.

\begin{table}
  \caption{Null and Alternate Hypothesis}
  \label{tab:hypothesis}
  \begin{tabular}{c}
    \toprule
    \(\mu\) = Mean of Differences \\
    \midrule
    \(H_0\) = \(_\mu\) App \(\le\) \(_\mu\) Book \\
    \midrule
    \(H_A\) = \(_\mu\) App > \(_\mu\) Book \\
    \midrule
    t-test result = P(2-tail)<0.05 hence \(H_0\) rejected\\
  \bottomrule
\end{tabular}
\end{table}

\begin{table}
  \caption{Summary of t-Test: Two-Sample
Assuming Unequal Variances (comparing the between-group-differences of the intervention groups based on the within-group-differences between pre-test and post-test scores). Learning through Printed brochure (Intervention Group 1 or IG 1). Learning through Quiz App (Intervention Group 2 or IG 2)}
  \label{tab:t-test}
  \begin{tabular}{ccl}
    \toprule
    &IG 2&IG 1\\
    \midrule
    Mean scores&2.64&1.84\\
    (post-test - pre-test)&&\\
    Variance&0.75&0.61\\
    Observations&78&76\\
    \midrule
    t-Stat&6.0\\
    P(T<=t) one-tail&0.000000007\\
    t-Critical one-tail&1.65\\
    P(T<=t) two-tail&0.000000014\\
    t-Critical two-tail&1.98\\
  \bottomrule
\end{tabular}
\end{table}

\section{Limitations}
The study only evaluated the changes in level of knowledge of Anganwadi Workers on the topic of Anemia. The impact of training interventions in terms of improvement of performance of AWWs could not be assessed. More studies and applications of game ideas or gamification systems and pedagogical strategies should be explored further and evaluated with larger groups of AWWs.

\section{Contributions}
Our contribution to Asian CHI research is by focusing towards upcoming smartphone user groups in developing countries with low resource, who need simple and effective learning methods to improve or refresh their knowledge base and skill-sets. In community healthcare and nursing, the quiz app can provide refresher training to the AWWs, enabling them to perform their duties better in the field.

\section{Conclusions}
Through our field experiment, we found that Android based quiz app was an effective tool for providing refresher training of AWWs. We designed an Android based quiz app in which AWWs played individually at their own pace and time. We observed that AWWs were able to show significant knowledge gains with knowledge retention. An improvement in ability to learn was observed in teaching the AWWs through the quiz based learning as compared to, teaching them by traditional text book method. Appropriate use of quiz as a pedagogy in the healthcare training curriculum of AWWs, resulted in well-prepared AWWs, with admirable knowledge about Anemia.

\begin{acks}
The authors would like to acknowledge all AWWs, AWW supervisors and CDPO Dharavi. Special thanks to Rucha Tulaskar for the illustrations, used in the app.
\end{acks}

\bibliographystyle{ACM-Reference-Format}
\bibliography{MendeleyReference}

@article{Perez2020,
    title = {{Development of an mHealth Behavior Change Communication Strategy: A case-study from rural Uttar Pradesh in India}},
    year = {2020},
    journal = {COMPASS 2020 - Proceedings of the 2020 3rd ACM SIGCAS Conference on Computing and Sustainable Societies},
    author = {P{\'{e}}rez, Myriam Cielo and Singh, Rohit and Chandra, Dinesh and Ridde, Valéry and Seth, Aaditeshwar and Johri, Mira},
    pages = {274--278},
    isbn = {9781450371292},
    doi = {10.1145/3378393.3402505}
}

@article{Nimmagadda2019,
    title = {{Effects of an mHealth intervention for community health workers on maternal and child nutrition and health service delivery in India: Protocol for a quasi-experimental mixed-methods evaluation}},
    year = {2019},
    journal = {BMJ Open},
    author = {Nimmagadda, Sneha and Gopalakrishnan, Lakshmi and Avula, Rasmi and Dhar, Diva and Diamond-Smith, Nadia and Fernald, Lia and Jain, Anoop and Mani, Sneha and Menon, Purnima and Nguyen, Phuong Hong and Park, Hannah and Patil, Sumeet R. and Singh, Prakarsh and Walker, Dilys},
    number = {3},
    pages = {1--10},
    volume = {9},
    doi = {10.1136/bmjopen-2018-025774},
    issn = {20446055},
    pmid = {30918034}
}

@article{WHO2016,
    title = {{Global strategy on human resources for health: Workforce 2030}},
    year = {2016},
    journal = {Who},
    author = {Organization, World Health},
    pages = {64},
    url = {https://www.who.int/hrh/resources/global_strategy_workforce2030_14_print.pdf?ua=1},
    isbn = {978 92 4 151113 1},
    issn = {1098-6596},
    pmid = {25246403},
    arxivId = {arXiv:1011.1669v3}
}

@article{Gragnolati2006,
    title = {{ICDS and persistent undernutrition: Strategies to enhance the impact}},
    year = {2006},
    journal = {Economic and Political Weekly},
    author = {Gragnolati, M and Bredenkamp, C and Gupta, M D and Lee, Y K and Shekar, M},
    number = {1201},
    volume = {1193}
}

@misc{ILA2021,
    title = {{ILA Module}},
    year = {2018},
    author = {{MoWCD}},
    url = {https://icds-wcd.nic.in/nnm/ILA.htm}
}

@article{Rao2018,
    title = {{India's integrated child development services scheme: challenges for scaling up}},
    year = {2018},
    journal = {Child: Care, Health and Development},
    author = {Rao, N. and Kaul, V.},
    number = {1},
    pages = {31--40},
    volume = {44},
    doi = {10.1111/cch.12531},
    issn = {13652214},
    pmid = {29235171}
}

@article{Chanani2016,
    title = {{M-Health for Improving Screening Accuracy of Acute Malnutrition in a Community-Based Management of Acute Malnutrition Program in Mumbai Informal Settlements}},
    year = {2016},
    journal = {Food and Nutrition Bulletin},
    author = {Chanani, Sheila and Wacksman, Jeremy and Deshmukh, Devika and Pantvaidya, Shanti and Fernandez, Armida and Jayaraman, Anuja},
    number = {4},
    pages = {504--516},
    volume = {37},
    doi = {10.1177/0379572116657241},
    issn = {03795721},
    pmid = {27370976}
}

@techreport{WHO2011,
    title = {{mHealth: New horizons for health through mobile technologies}},
    year = {2011},
    booktitle = {Observatory},
    author = {{WHO}},
    pages = {102},
    volume = {3},
    url = {http://www.who.int/goe/publications/goe_mhealth_web.pdf%5Cnhttp://www.who.int/goe/publications/ehealth_series_vol3/en/index.html},
    institution = {World Health Organization},
    address = {Geneva}
}

@article{Kumar2015,
    title = {{Mobile phones for maternal health in rural India}},
    year = {2015},
    journal = {Conference on Human Factors in Computing Systems - Proceedings},
    author = {Kumar, Neha and Anderson, Richard},
    pages = {427--436},
    volume = {2015-April},
    isbn = {9781450331456},
    doi = {10.1145/2702123.2702258}
}

@article{Ramachandran2010,
    title = {{Mobile-izing health workers in rural India}},
    year = {2010},
    journal = {Conference on Human Factors in Computing Systems - Proceedings},
    author = {Ramachandran, Divya and Canny, John and Das, Prabhu Dutta and Cutrell, Edward},
    pages = {1889--1898},
    volume = {3},
    isbn = {9781605589299},
    doi = {10.1145/1753326.1753610}
}

@inproceedings{Shah2017,
    title = {{Tackling child malnutrition: An innovative approach for training health workers using ICT: A pilot study}},
    year = {2017},
    booktitle = {IEEE Region 10 Humanitarian Technology Conference 2016, R10-HTC 2016 - Proceedings},
    author = {Shah, Mithilesh P. and Kamble, Pawan A. and Agnihotri, Satish B.},
    isbn = {9781509041770},
    doi = {10.1109/R10-HTC.2016.7906811}
}

\appendix
\section{Screenshots of mobile quiz app}

\begin{figure}
\centering
\begin{tabular}{ll}
\includegraphics[width=0.432\linewidth]{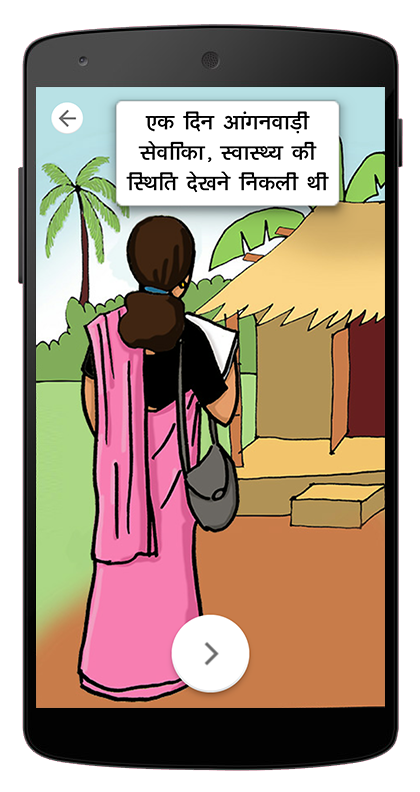}
&
\includegraphics[width=0.432\linewidth]{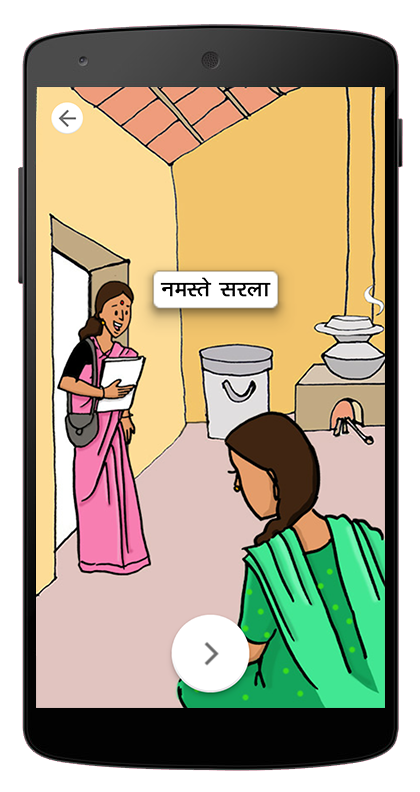}\\

\includegraphics[width=0.432\linewidth]{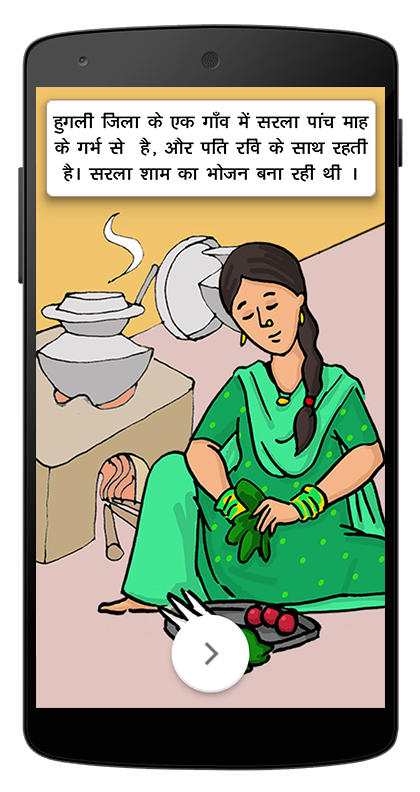}
&
\includegraphics[width=0.432\linewidth]{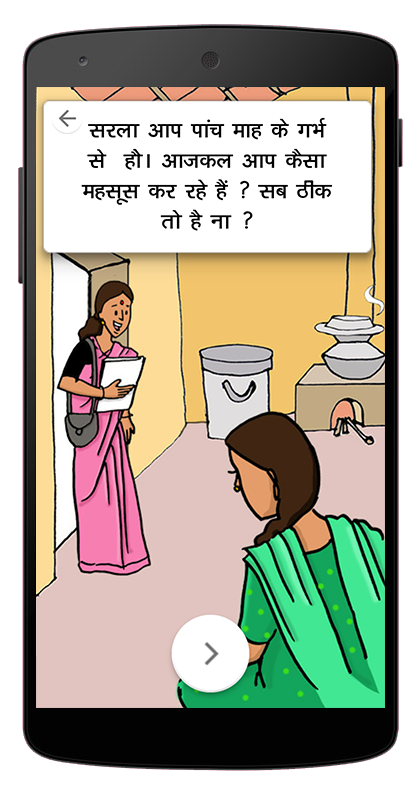}\\

\includegraphics[width=0.432\linewidth]{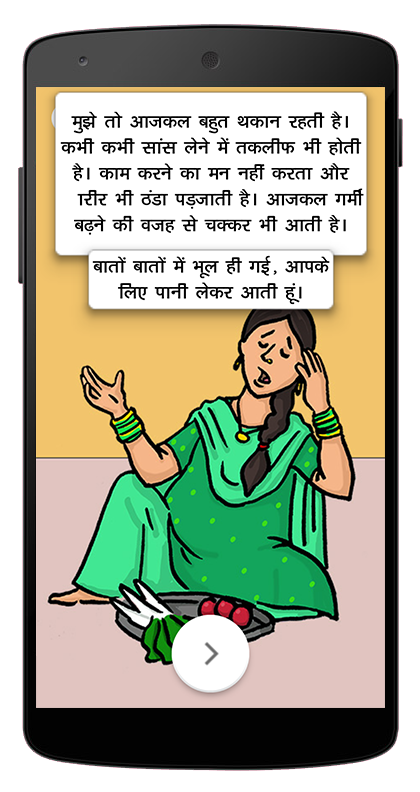}
&
\includegraphics[width=0.432\linewidth]{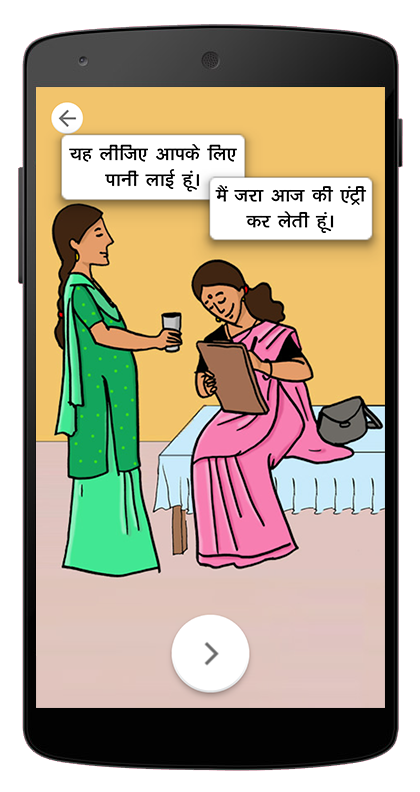}
\end{tabular}
\caption{Screenshots of the quiz app}
\Description{Screenshots of the quiz app}
\end{figure}

\begin{figure}
\centering
\begin{tabular}{ll}
\includegraphics[width=0.432\linewidth]{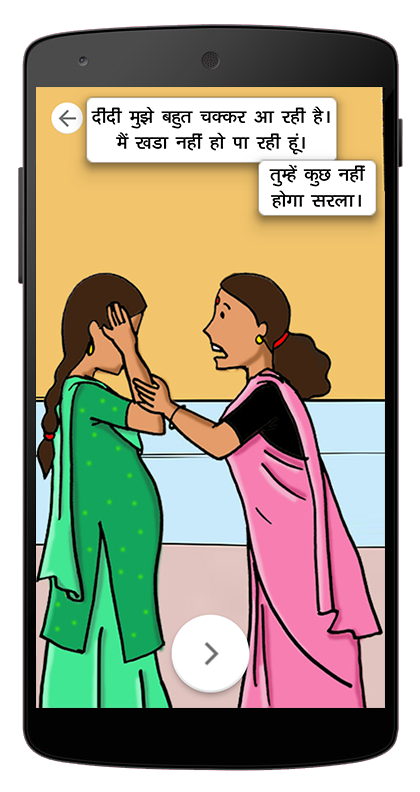}
&
\includegraphics[width=0.432\linewidth]{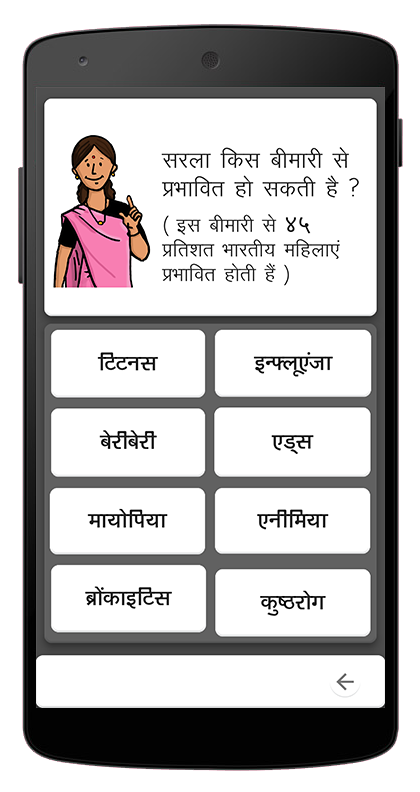}
\\
\includegraphics[width=0.432\linewidth]{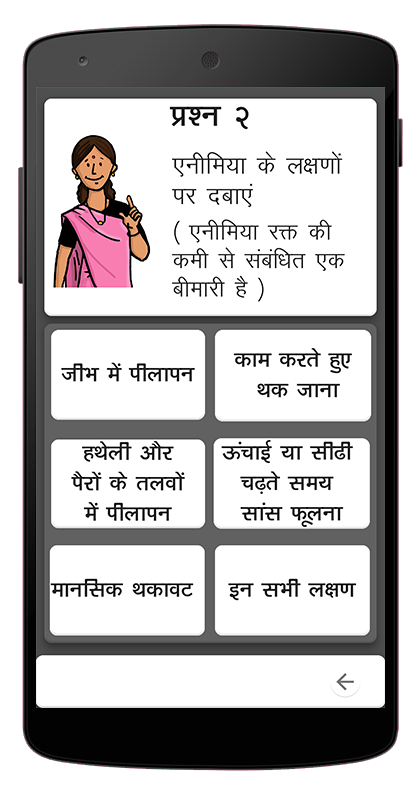}
&
\includegraphics[width=0.432\linewidth]{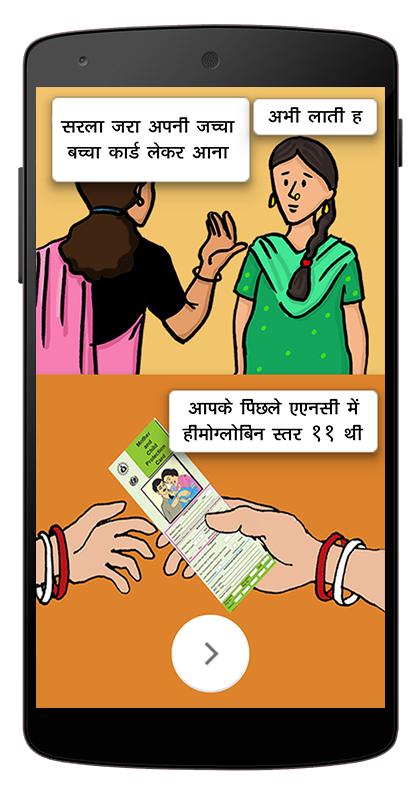}
\\
\includegraphics[width=0.432\linewidth]{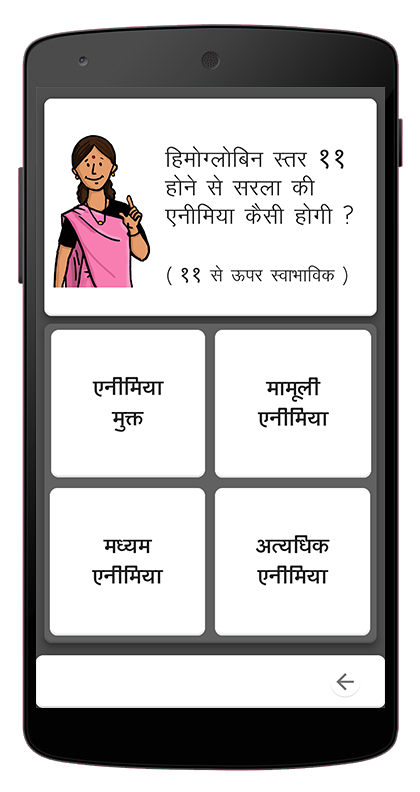}
&
\includegraphics[width=0.432\linewidth]{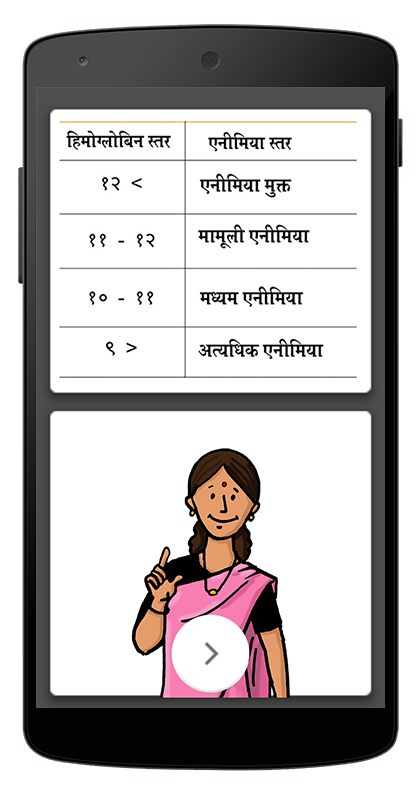}
\end{tabular}
\caption{Screenshots of the quiz app}
\Description{Screenshots of the quiz app}
\end{figure}

\begin{figure}
\centering
\begin{tabular}{ll}
\includegraphics[width=0.432\linewidth]{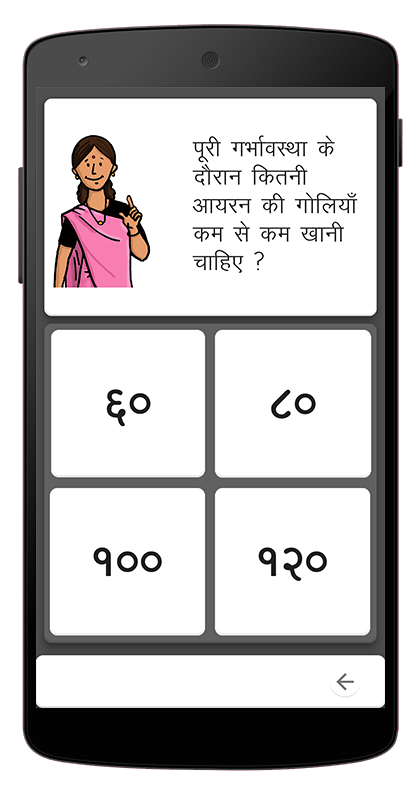}
&
\includegraphics[width=0.432\linewidth]{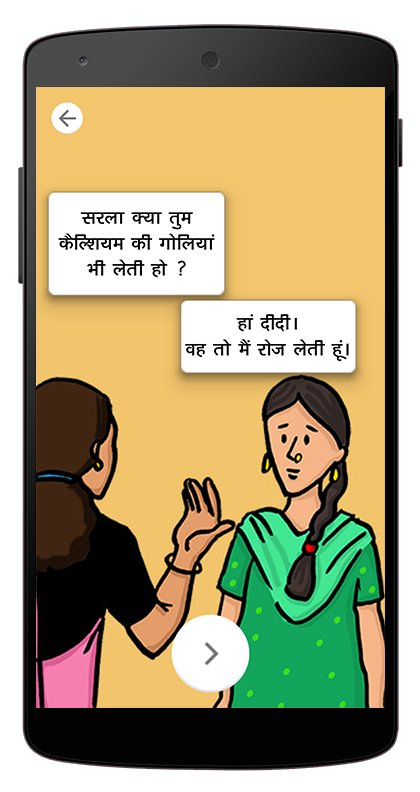}
\\
\includegraphics[width=0.432\linewidth]{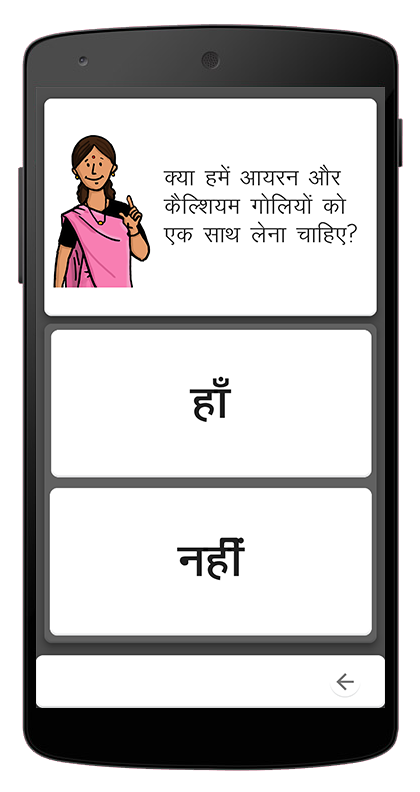}
&
\includegraphics[width=0.432\linewidth]{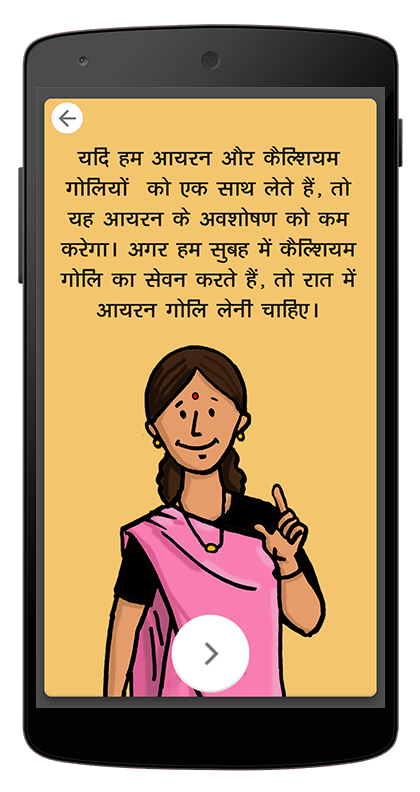}
\\
\includegraphics[width=0.432\linewidth]{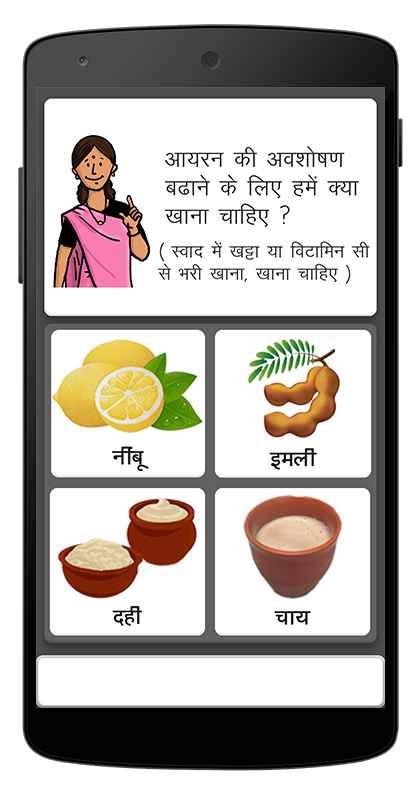}
&
\includegraphics[width=0.432\linewidth]{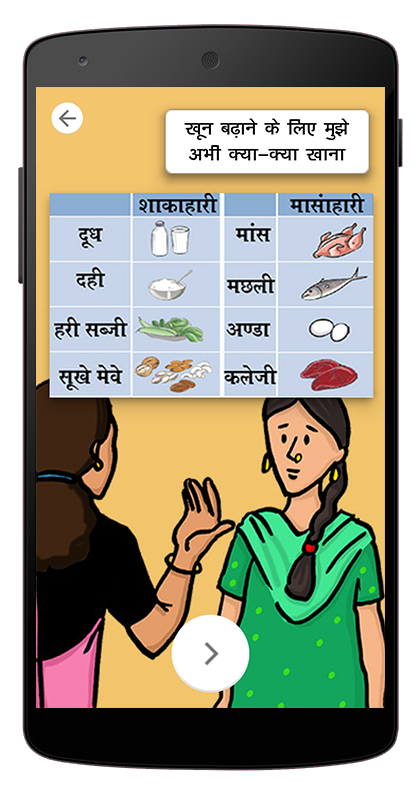}
\end{tabular}
\caption{Screenshots of the quiz app}
\Description{Screenshots of the quiz app}
\end{figure}

\begin{figure}
\centering
\begin{tabular}{ll}
\includegraphics[width=0.432\linewidth]{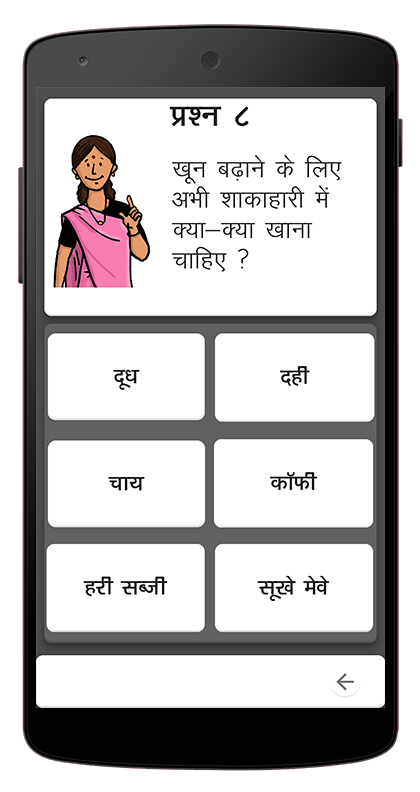}
&
\includegraphics[width=0.432\linewidth]{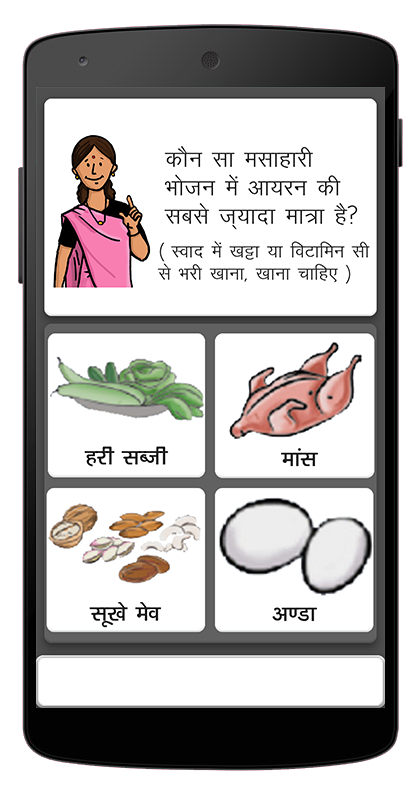}
\\
\includegraphics[width=0.432\linewidth]{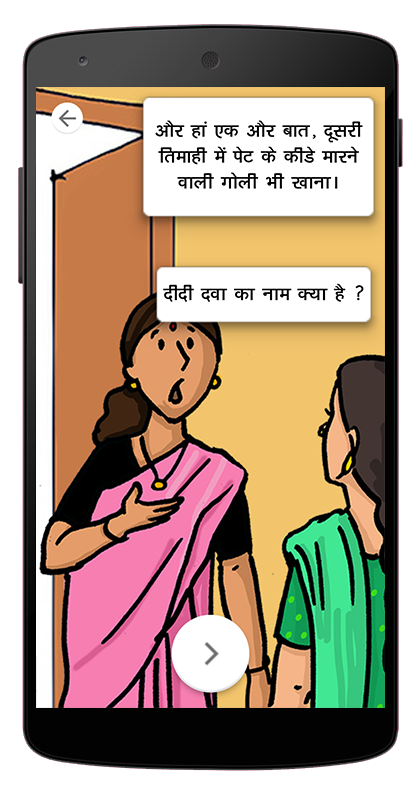}
&
\includegraphics[width=0.432\linewidth]{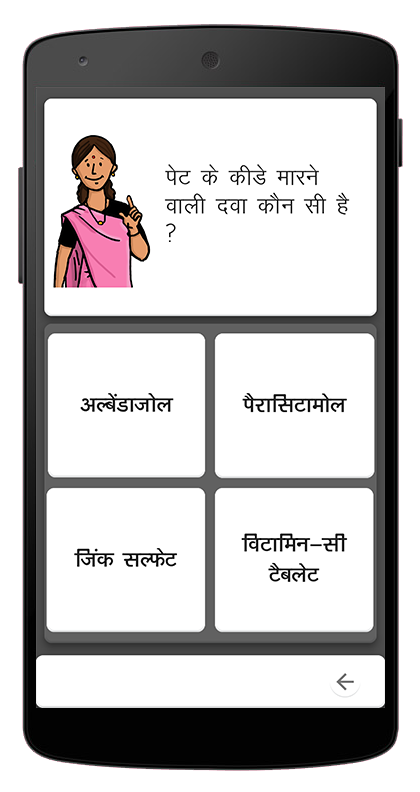}
\\
\includegraphics[width=0.432\linewidth]{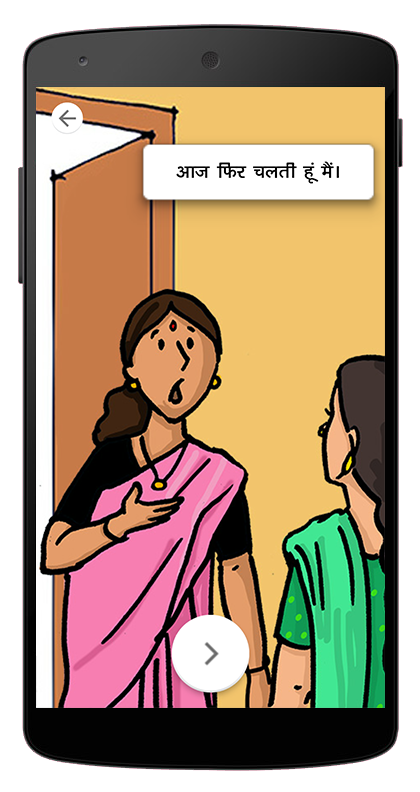}
&
\includegraphics[width=0.4\linewidth]{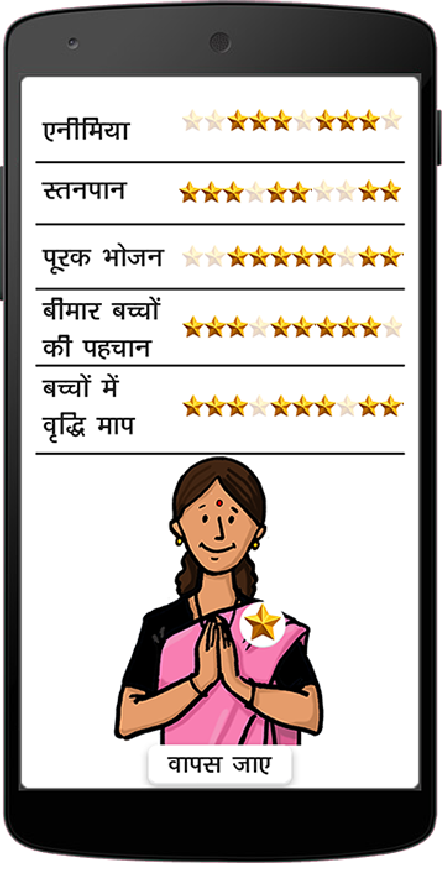}
\end{tabular}
\caption{Screenshots of the quiz app}
\Description{Screenshots of the quiz app}
\end{figure}

\newpage

\section{Takeaway ILA Module - Anemia}
\begin{figure}
\centering
\begin{tabular}{l}
\\
\\
\includegraphics[width=0.85\textwidth]{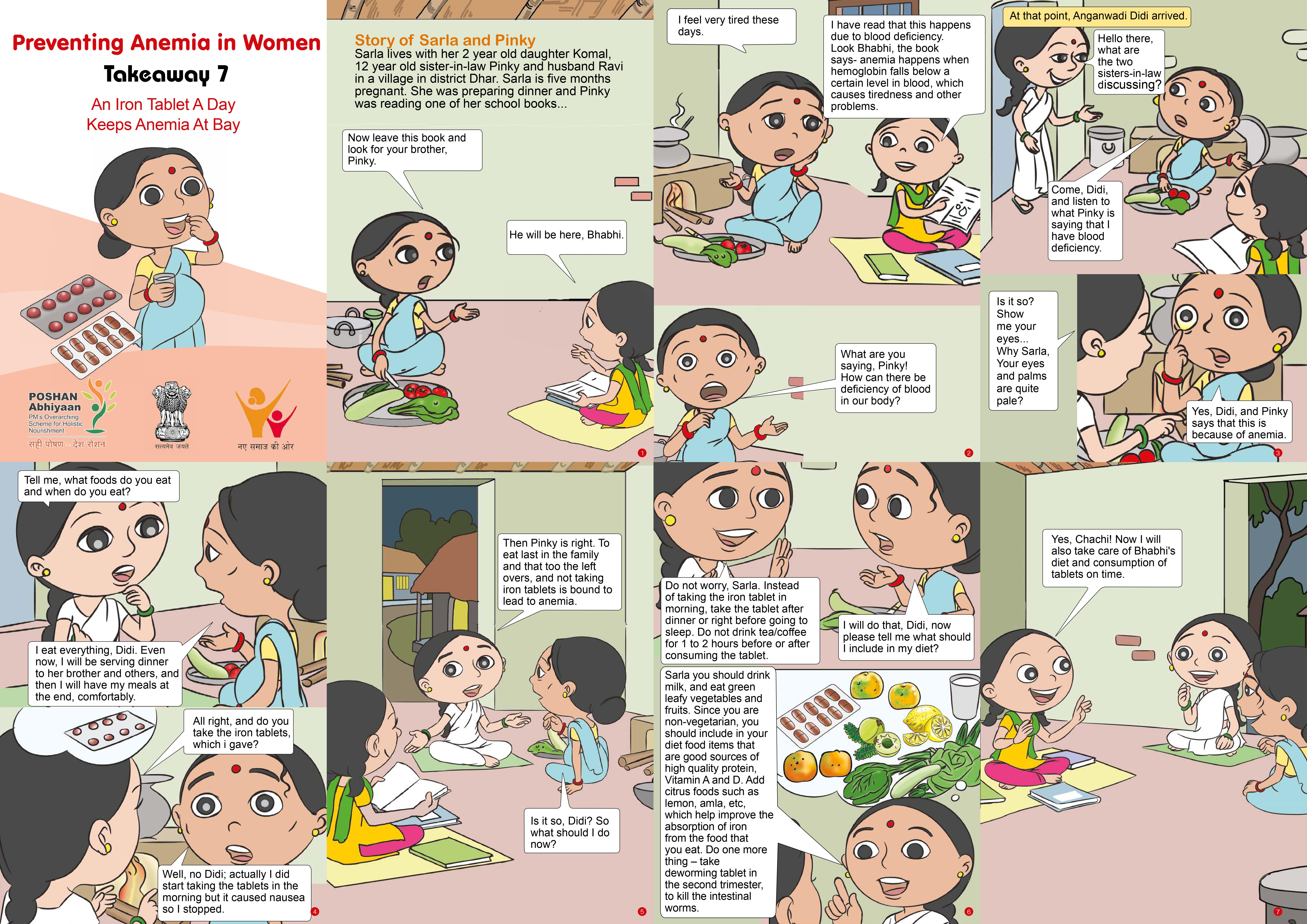}\\
\includegraphics[width=0.85\textwidth]{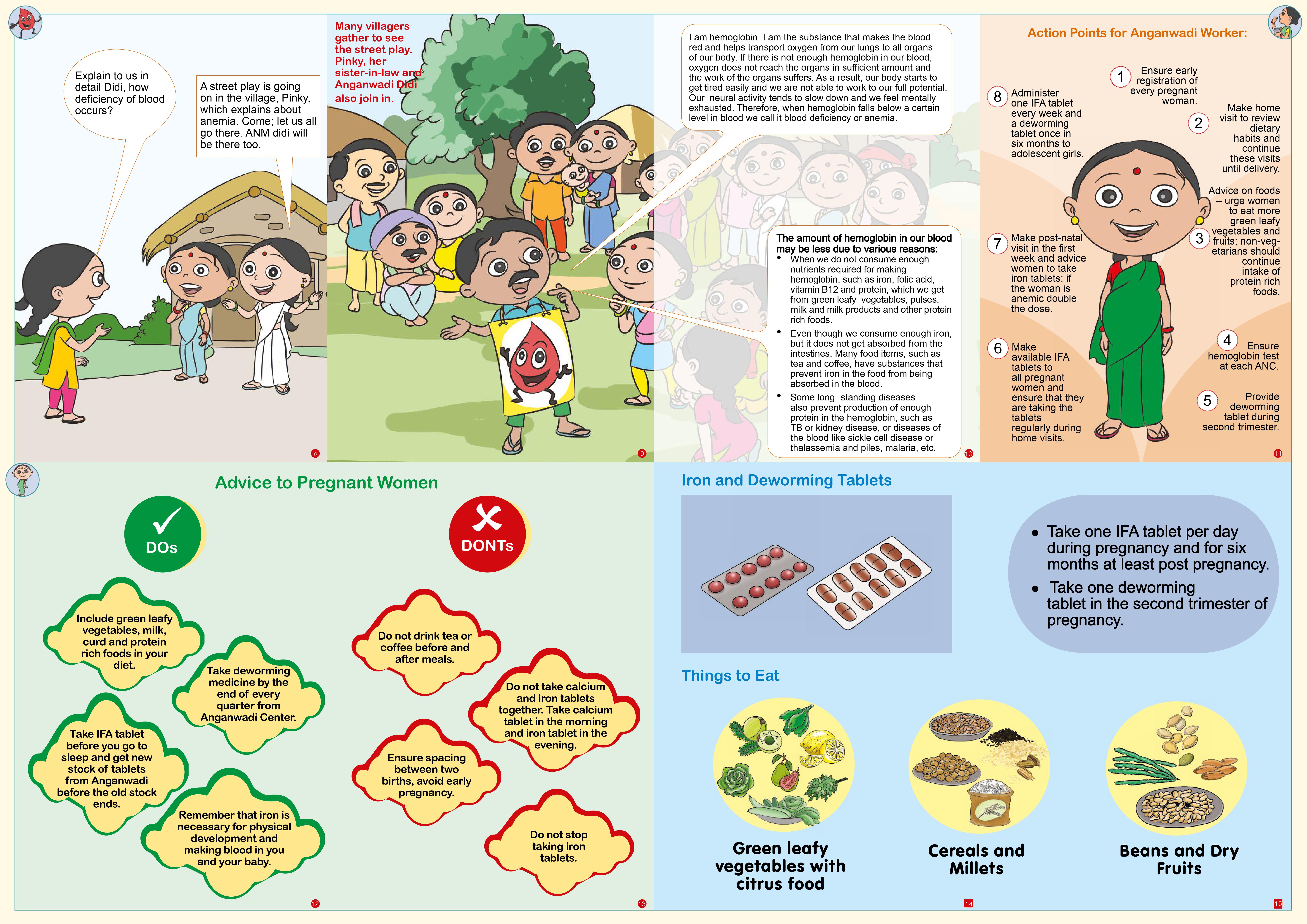}
\end{tabular}
\caption{Takeaway ILA Module - Anemia, Page 1 and 2}
\Description{Takeaway ILA Module - Anemia, Page 1 and 2}
\end{figure}

\end{document}